\documentclass[preprint2]{aastex}

\begin{document}
\title{Ionized Gas in the Irr Galaxy~IC~10: The Emission Spectrum
and Ionization Sources}

\author{V.P. Arkhipova \altaffilmark{1}, O.V. Egorov \altaffilmark{1}, T.A. Lozinskaya \altaffilmark{1},
 A.V. Moiseev \altaffilmark{2}}

\altaffiltext{1}{Sternberg Astronomical Institute,
Universitetskiy pr. 13, Moscow, 119992 Russia, E-mail: vera@sai.msu.ru}

\altaffiltext{2}{Special Astrophysical Observatory, Russian
Academy of Sciences, Nizhniy Arkhyz, Karachai-Cherkessian
Republic, 357147 Russia}

\begin{abstract}
We present the results of observations of the Irr galaxy~IC~10 at
the 6-m SAO telescope with the panoramic Multi-Pupil Fiber
Spectrograph (MPFS). Based on the results of these observations and
our long-slit spectroscopy performed previously, we have
investigated the ionized-gas emission spectrum in the region of
intense star formation and refined the gas metallicity estimates. We
show that the ``diagnostic diagrams'' constructed from our
observations agree best with the new improved ionization models by
Mart\'in-Manj\'on et~al. Using these models, we have determined the
electron density and gas ionization parameter and ionizing-cluster
characteristics, the age and mass, from the spectra of the
investigated HII~regions. The cluster age and mass are shown to be
within the ranges 2.5 -- 5~Myr and (0.2 -- 1)$\times10^{5}
M_{\odot}$, respectively.
\end{abstract}


\maketitle

\section*{INTRODUCTION}

Studies of Local Group irregular galaxies open optimal possibilities for investigating
the structure, kinematics, and chemical composition of the gaseous medium and their
changes not only during the evolution of Irr~galaxies but, according to present views,
also during the evolution of massive galaxies. The topicality of detailed
studies of the gas emission spectrum and ionization conditions in the nearest dwarf
Irr~galaxy IC~10 is determined by a number of its peculiarities. The anomalously large
number of Wolf--Rayet (WR) stars in IC~10 arouses the greatest interest. The space
density of WR~stars here is highest among the dwarf galaxies, comparable to that in
massive spiral galaxies (Massey et~al.~1992; Richer et~al.~2001; Massey and
Holmes~2002; Crowther et~al.~2003; Vacca et~al.~2007). The stellar population of~IC~10
is indicative of both a recent starburst (t=3 -- 10~Myr) and an older starburst
($t>350$~Myr) (Hunter~2001; Zucker~2002; Massey et~al.~2007; Vacca et~al.~2007; and
references therein). The high H$\alpha$ and infrared luminosities and the anomalously
large number of WR~stars in IC~10 suggest that the last starburst was short but it
affected the bulk of the galaxy.

The gas and stellar compositions of the galaxy have been well
studied (and are being intensively studied). Not coincidentally, it
is for~IC~10 that Yin et~al.~(2010) performed detailed model
calculations of the evolution of the chemical composition in an
Irr~galaxy for various star formation and galactic
wind regimes.

In this paper, we continue our studies of the gas emission spectrum
in the dwarf Irr starburst galaxy~IC~10 begun by Lozinskaya
et~al.~(2009). The main objective of the first paper was to
determine the gas metallicity in about twenty HII~regions and in the
synchrotron suberbubble from our observations of the galaxy at the
6-m SAO telescope with the SCORPIO focal reducer in the long-slit
mode (the chemical composition of only the three brightest nebulae
in the galaxy was known previously; see Lozinskaya et~al.~(2009) and
references therein.) The paper by Magrini and Gon\c{c}alves~(2009),
who determined the chemical composition of many HII~regions and
planetary nebulae in~IC~10, appeared in the same year.

The main goal of our new paper is to analyze the emission spectrum
and gas ionization state by the galaxy's stellar
population. This study is based on our observations performed at
the 6-m SAO telescope with the panoramic Multi-Pupil Fiber
Spectrograph (MPFS). Since these observations are interpreted in
terms of the new improved ionization models by Mart\'in-Manj\'on
et~al.~(2009) and are compared with the new computations by
Levesque et~al.~(2009), we deemed it appropriate to also
reconsider the results of our long-slit spectroscopy presented in
Lozinskaya et~al.~(2009) in terms of these models. The data on the
stellar population of~IC~10 were also updated significantly over
the year: two papers devoted to the search for star clusters in
the galaxy appeared simultaneously (Tikhonov and
Galazutdinova~2009; Sharina et~al.~2009). Previously,
Hunter~(2001) identified clusters only in the region of intense
current star formation and in its immediate neighborhood; in the
mentioned new papers, the investigated galactic regions were
extended and the number of clusters increased significantly.

The most interesting region that we studied in greatest detail is the bright complex of
ionized gas in the southeastern section of the galaxy, in which the last starburst
episode most likely occurred in~IC~10 (Vacca et~al.~2007). Here, there are the densest
HI~cloud, a molecular CO~cloud, and a complex of emission nebulae about 300--400~pc in
size, including two bright shell nebulae, HL111 and~HL106 (according to the catalog by
Hodge and Lee~(1990)), as well as the youngest star clusters and ten WR~stars (Leroy
et~al.~2006; Wilcots and Miller~1998; Gil de Paz et~al.~2003; Lozinskaya et~al.~2009;
Egorov et~al.~2010; and references therein). The size of the shell~HL111 determined by
its three arc parts~HL111c, HL111d, and~HL111e is about~$10''$ or 39~pc. An object
previously identified as the WR~star~M24 (in what follows, the WR~stars from the lists
by Royer et~al.~(2001) and Massey and Holmes~(2002) are denoted by the letters~R and~M,
respectively) is located in the nebula HL111c, which represents the brightest part of
the shell. According to present views (see Vacca et~al.~2007), M24 is a close group of
stars that consists of at least six blue stars; four of them are possible
WR~candidates.

The galaxy's H$\alpha$ and~[SII] images reveal fainter multiple
shells and supershells with sizes from 50~pc to 800--1000~pc outside
this brightest complex (Zucker~2000; Wilcots and Miller~1998; Gil de
Paz et~al.~2003; Leroy et~al.~2006; Chyzy et~al.~2003; Lozinskaya
et~al.~2008). Some of the MPFS fields and long-slit spectrograms are
localized in these faint regions.

A unique object, the so-called synchrotron suberbubble, is adjacent
to the central bright star formation complex on the south.
Lozinskaya and Moiseev~(2007) were the first to explain the
formation of the synchrotron superbubble by a hypernova explosion;
previously, its formation was associated with multiple explosions of
about ten supernovae (Yang and Skillman~1993; Bullejos and
Rozado~2002; Rosado et~al.~2002; Thurow and Wilcots~2005).

In the succeeding sections, we describe our observations, present
and discuss our results, and, in conclusion, summarize our main
conclusions.

All radial velocities here are heliocentric; the distance to the
galaxy is taken be 800~kpc (the angular scale is $\simeq$
3.9~pc$/''$) (Sanna et~al.~2009; Tikhonov and Galazudtdunova~2010).

\begin{figure*}[t!]

\center{~~~~~~~~~~~(a)\hfill

\includegraphics[width=0.8\linewidth]{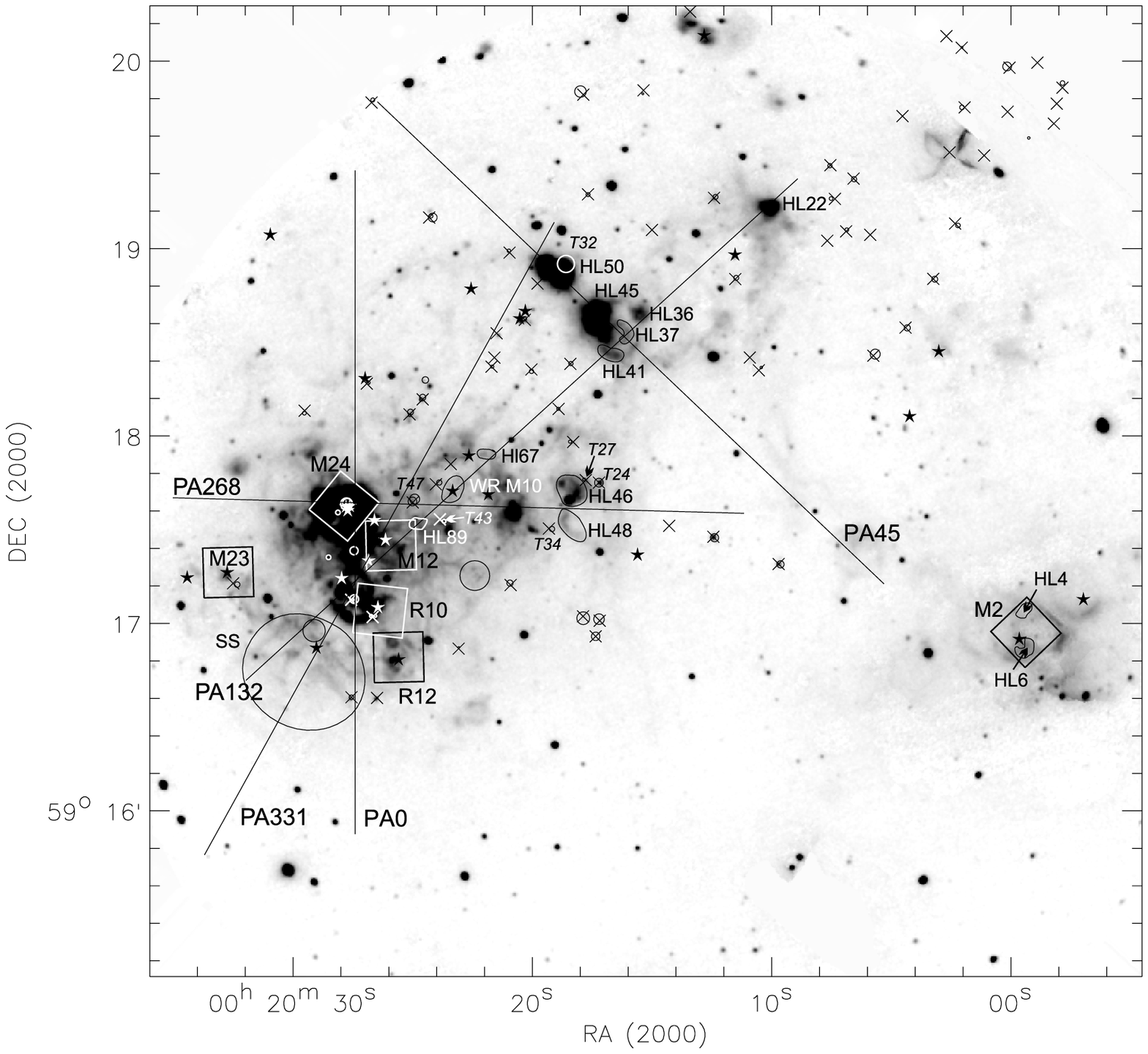}
}
\caption{Localization of five long-slit spectrograms
and six MPFS fields in the H$\alpha$ image of IC~10: (a)~the entire
galaxy and (b)~the brightest region of current star formation. The
MPFS fields are designated according to their central WR~star; the
long-slit spectrograms, as in Lozinskaya et~al.~(2009), are
designated according to their position angle. The asterisks mark the
spectroscopically confirmed WR~stars from the lists by Royer
et~al.~(2001) and Massey and Holmes~(2002). The circles indicate the
star clusters from the lists by Hunter~(2001) and Tikhonov and
Galazutdinova~(2010); the crosses indicate the centers of the
clusters from the list by Sharina et~al.~(2009). The clusters used
in the text are denoted by the letter~T and the number from the list
by Tikhonov and Galazutdinova~(2010). Also shown are the HII~regions
over which the lines presented in Table~3 were integrated.\hfill}\label{fig:loc-all}
\end{figure*}

\begin{figure*}
\center{~~~~~~~~~~(b)\hfill

\includegraphics[width=0.75\linewidth]{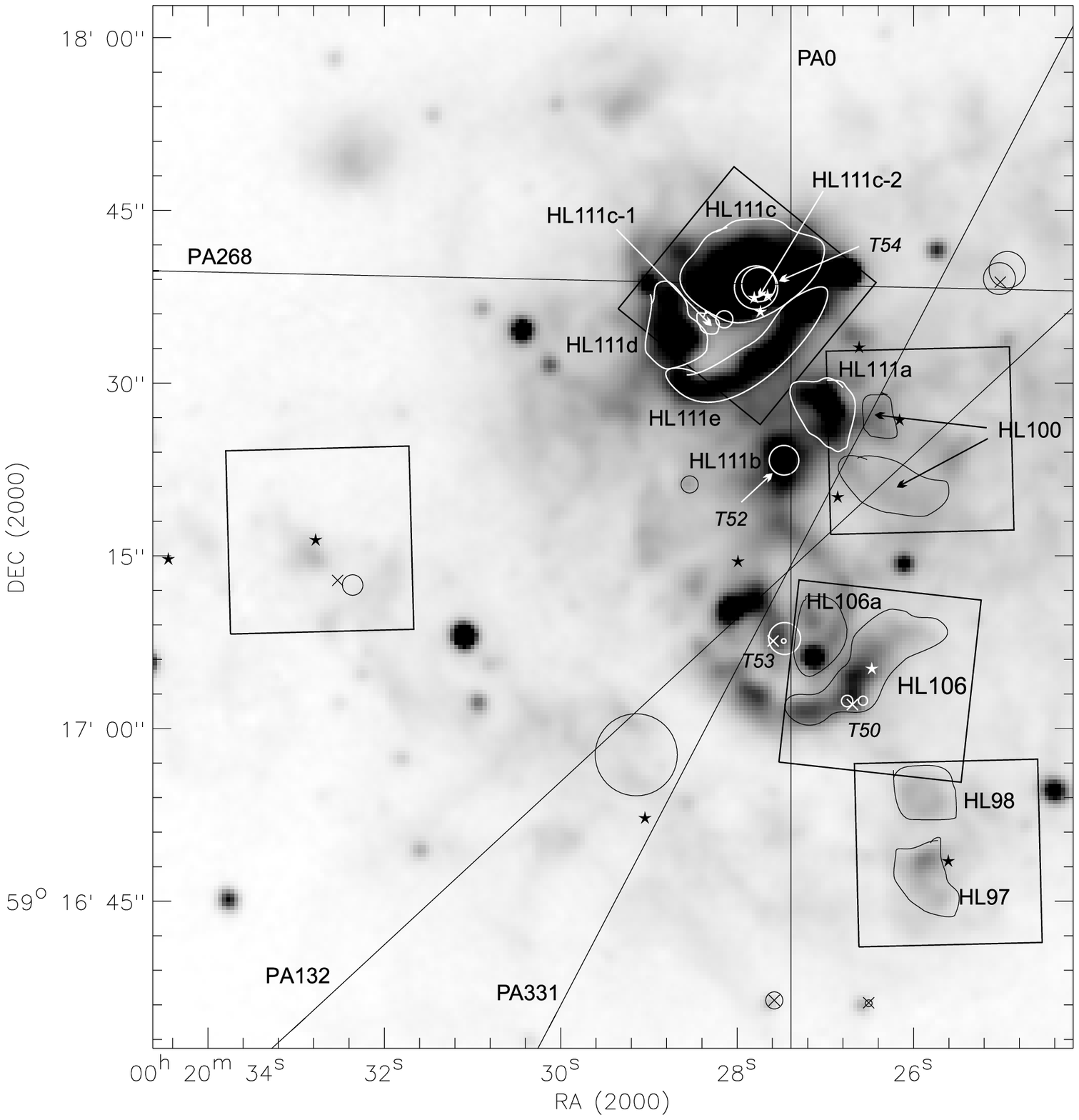}}
\addtocounter{figure}{-1}
\caption{(Contd.)}\label{fig:loc-m24}
\end{figure*}

\section*{OBSERVATIONS AND DATA REDUCTION}

\subsection*{MPFS Observations}

The selected galactic regions were observed with the panoramic Multi-Pupil Fiber
Spectrograph (MPFS) mounted at the prime focus of the 6-m telescope (see Afanasiev
et~al.~(2001) and the web site\footnote{http://www.sao.ru/hq/lsfvo/devices.html}). The detector
was an EEV 42--20 $2048\times2048$-pixel CCD~array. The spectrograph can simultaneously
record the spectra from 256 spatial elements (spaxels) (in the shape of square lenses)
that constitute a $16\times16$-spaxel array in the plane of the sky. The angular size
of a single spaxel is~$1''$.

In this paper, we use the spectra of five fields with a resolution
of about 6.5~\AA\ in the wavelength range 3990--6940~\AA\ taken
for~IC~10 (in the sixth field around the WR star M23, the emission
turned out to be very weak.) The localization of the MPFS fields in
the galactic image is shown in Fig.~1; the fields are designated
according to the central WR~star.

We reduced the observations using the software developed at the
SFVO laboratory of the SAO, the Russian Academy of Sciences, and
running in the IDL environment. The spectra of the stars
BD+75$^{d}$325 and G191-B2B observed immediately after the object
were used for energy calibration. The result of this reduction is
a ``data cube'' in which a 2048-pixel spectrum corresponds to each
spaxel of the $16''\times16''$ image.

\subsection*{Long-Slit Spectroscopy}

We also used the observations of IC~10 with the SCORPIO instrument
(for a description, see Afanasiev and Moiseev~2005) operating in the
long-slit mode. The observations were performed with a slit
about~$6'$ in length and~$1''$ in width; the scale along the slit
was~$0.36''$ per pixel. The technique of these observations and the
results obtained are described in detail in Lozinskaya~et~al.~(2009).

For five slit positions designated in accordance with their
position angles as PA0, PA45, PA132, PA268, and PA331, we took
spectra with a resolution from~2.5 to 9~\AA. The localization of
the spectrograms is shown in Fig.~1.

To increase the signal-to-noise ratio for faint emission regions, we
performed an averaging over individual nebulae when processing the
spectrograms; the region of integration was from~2 to~$20''$ in
size.

The ranges of errors given below in the tables and figures
correspond to~$3\sigma$.

A log of MPFS and SCORPIO long-slit observations is presented in Table~1. Its columns
give: (1)--- MPFS field designated according to the central WR~star or spectrogram
designated according to its position angle; (2)--- date of observations; (3)---
spectral range; (4)--- spectral resolution; (5)--- total exposure time; (6)---
``seeing''.

Here, we also partly use the observational data obtained with
SCORPIO in the mode of a Fabry\mbox{--}Perot interferometer in
the~H$\alpha$ and~[SII] lines presented previously in Egorov
et~al.~(2010).

\begin{table*}[p!]
\caption{Log of observations}
\center{\begin{tabular}{|l|c|c|c|c|c|}
\hline MPFS field/spectrum (PA) & Date        &
Range  $\Delta\lambda$, \AA& $\delta\lambda$, \AA & $T_{\exp}$, s & seeing \\
 \hline
MPFS M24 & Aug.~8/9, 2004& 3990--6940 & 6.5  &2700 & 1.7--2.0 \\
MPFS R10 & Aug.~8/9, 2004& 3990--6940 & 6.5  &1800 & 1.4 \\
MPFS M12 & Sep.~28/29, 2005& 3990--6940 & 6.5  &3600 & 2.0 \\
MPFS R12 & Sep.~28/29, 2005& 3990--6940 & 6.5  &3600 & 2.0 \\
MPFS M2  & Aug.~8/9, 2004& 3990--6940 & 6.5  &1200 & 1.5 \\
MPFS M23 & Sep.~28/29, 2005& 3990--6940 & 6.5  &2400 & 2.5 \\
\hline
PA0      & Aug.~17/18, 2006& 3620--5370 & 5\phantom{.0}  &6000 & 2.5 \\
PA0      & Sep.~2/3, 2008& 5650--7340 & 5\phantom{.0}  &3600 & 2.1 \\
PA132    & Aug.~17/18, 2006& 3620--5370 & 5\phantom{.0}  &4800 & 1.4 \\
PA132    & Aug.~18/19 2006& 5650--7340 & 5\phantom{.0}  &3600 & 1.4 \\
PA268    & Feb.~14/15, 2007& 6060--7060 & 2.5&1200 & 1.4 \\
PA268    & Jan.~15/16, 2008& 3620--5370 & 5\phantom{.0}  &4800 & 1.5 \\
PA45     & Oct.~29/30, 2008& 3650--7350 & 9\phantom{.0}  &3600 & 1.0 \\
PA331    & Oct.~28/29, 2008& 3650--7350 & 9\phantom{.0}  &6000 & 0.9 \\
\hline
\end{tabular}}
\end{table*}

\begin{table*}[p!]
%%% Table:2
\caption{Relative line intensities in the regions of intersection
between the spectrograph slits and MPFS fields}
\center{\begin{tabular}{|l|c|c|c|}
\hline \multicolumn{1}{|p{0.1\linewidth}|}{MPFS and spectrum} &
$\log$($I$([OIII]4959,5007)/$I$(H$\beta$)) &
$\log$($I$([SII]6717,31)/$I$(H$\alpha$)) &
$\log$($I$([NII]6584)/$I$(H$\alpha$))\\
\hline
M24   & $0.69 \pm 0.02$ & $-0.97 \pm 0.07$ & $-1.31 \pm 0.05$\\
PA268 & $0.66 \pm 0.02$ & $-0.97 \pm 0.02$ & $-1.33 \pm 0.02$\\
\hline
M24   & $0.70 \pm 0.02$ & $-0.98 \pm 0.07$ & $-1.27 \pm 0.04$\\
PA0 & $0.67 \pm 0.03$ & $-0.86 \pm 0.02$ & $-1.32 \pm 0.02$\\
\hline
M12   & $0.47 \pm 0.08$ & $-0.32 \pm 0.09$ & $-0.84 \pm 0.08$\\
PA331 & $0.45 \pm 0.07$ & $-0.42 \pm 0.02$ & $-1.00 \pm 0.05$\\
\hline
M12   & $0.41 \pm 0.08$ & $-0.21 \pm 0.07$ & $-0.84 \pm 0.09$\\
PA132 & $0.34 \pm 0.09$ & $-0.18 \pm 0.02$ & $-0.75 \pm 0.04$\\
\hline
R10  & $0.43 \pm 0.13$ & $-0.42 \pm 0.09$ & $-0.93 \pm 0.05$ \\
PA0 & $0.58 \pm 0.11$ & $-0.57 \pm 0.05$ & $-0.93 \pm 0.03$ \\
\hline
\end{tabular}}
\end{table*}

\begin{table*}[t!]
%%% Table:3

\caption{Relative line intensities from our MPFS observations}
\center{\begin{tabular}{|l|c|c|c|}
\hline \multicolumn{1}{|c|}{MPFS field and HII~region} &
$I$([OIII]4959,5007)/$I$(H$\beta$) &
$I$([SII]6717,31)/$I$(H$\alpha$) &
$I$([NII]6584)/$I$(H$\alpha$)\\
\hline
M24 HL111c   & $4.74 \pm 0.39$ & $0.123 \pm 0.038$ & $0.058 \pm 0.015$\\
M24 HL111c-1 & $5.13 \pm 0.44$ & $0.131 \pm 0.027$ & $0.058 \pm 0.011$\\
M24 HL111c-2 & $5.32 \pm 0.14$ & $0.087 \pm 0.009$ & $0.063 \pm 0.025$\\
M24 HL111d   & $4.91 \pm 0.62$ & $0.156 \pm 0.036$ & $0.066 \pm 0.016$\\
M24 HL111e   & $3.98 \pm 0.40$ & $0.212 \pm 0.063$ & $0.083 \pm 0.015$\\
\hline
M12 HL100 & $1.93 \pm 0.57$ & $0.66 \pm 0.14$ & $0.162 \pm 0.022$\\
M12 HL111a & $4.00 \pm 1.50$ & $0.38 \pm 0.09$ & $0.123 \pm 0.023$\\
\hline
R10  HL106 & $2.49 \pm 0.73$ & $0.39 \pm 0.09$ & $0.129 \pm 0.004$\\
R10  HL106a & $2.89 \pm 0.69$ & $0.25 \pm 0.04$ & $0.102 \pm 0.018$\\
\hline
R12 HL97 & $5.8 \pm 2.4 $ & $0.52 \pm 0.12$ & $0.153 \pm 0.044$\\
R12 HL98 & $3.4 \pm 1.2$ & $0.75 \pm 0.25$ & $0.165 \pm 0.033$\\
\hline
M2 HL4 & $4.6  \pm 3.6$ & $0.73 \pm 0.22$ & $0.162 \pm 0.128$\\
M2 HL6 & $8.4  \pm 5.8$ & $0.40 \pm 0.18$ & $0.117 \pm 0.046$\\
\hline
\end{tabular}}
\end{table*}

\section*{RESULTS OF OBSERVATIONS}

The localization of the six MPFS fields and five long-slit
spectrograms that were analyzed in detail in Lozinskaya
et~al.~(2009) and partially used in this paper in the H$\alpha$
image of the galaxy is shown in Fig.~1. The spectroscopically
confirmed WR~stars from the lists by Royer et~al.~(2001) and Massey
and Holmes~(2002) as well as the star clusters from the lists by
Hunter~(2001), Tikhonov and Galazutdinov~(2010), and Sharina
et~al.~(2010) are labeled in the figure.

In the list of cluster coordinates presented by Tikhonov and
Galazutdinova~(2010) and in the coordinates of the M24 group of
stars in Vacca et~al.~(2007), we revealed a shift by about~$2''.4$
related to the referencing of the coordinate system of HST
observations. After an appropriate refinement of the coordinates
(Vacca, private communication; Galazutdinova, private
communication), the localizations of the common clusters in the
three mentioned lists in Fig.~1 agree well in most cases (the
refined coordinates are given in the updated version of the paper
by Tikhonov and Galazutdinova (2009): arXiv:1002.2046v1
[astro-ph.GA])

For the convenience of identification, the HII~regions from the list
by Hodge and Lee~(1990) mentioned in the text and the synchrotron
superbubble (denoted by~SS in Fig.~1a) are also labeled in Figs.~1a
and~1b. The two separate areas of the nebula HL111c in Fig.~1b and
in Table~3 (see below) are designated as HL111c-1 and HL111c-2.

The observations of several galactic regions were performed both with the MPFS and
long-slit spectrograph (see Fig.~1), which allows the actual accuracy of our
measurements to be estimated. For this purpose, we determined the integrated relative
intensities of several lines for the regions falling into the spectrograph slit and cut
out the same regions in the corresponding MPFS~field. Table~2 gives the relative line
intensities in the same region measured from our long-slit spectrograms and MPFS
observations. As we see, the agreement is good everywhere, within the error limits.

Previously (Lozinskaya et~al.~2009), we made sure that the measurements made from two
long-slit spectra at the points of their intersection are also in good agreement, given
the difference between the regions of integration of the fluxes for different slit
orientations.

The results of our processing of the MPFS~spectra are presented in Table~3. Its columns
give: (1)--- MPFS field used and name of the HII~region or its part over which the line
flux was integrated; (2), (3), and (4) --- integrated relative line intensities
$I$([OIII]4959,5007)/$I$(H$\beta$), $I$([SII]6717,31)/$I$(H$\alpha$), and
$I$([NII]6584)/$I$(H$\alpha$), respectively. The regions over which the integration was
performed are shown in Figs.~1a  and~1b; the two parts of the bright nebula HL111c are
marked by the indices~1 and~2.

The relative line intensities averaged over the long-slit
spectroscopic observations of the investigated nebulae are listed in
Table~3 from Lozinskaya et~al.~(2009).

As an example, Fig~2 shows the distribution maps of relative line intensities:
$I$([SII]$\lambda6717+6731$\AA)\slash $I$(H$\alpha$),
$I$([NII]$\lambda6584$\AA)\slash $I$(H$\alpha$),
 and \linebreak $I$([OIII]$\lambda$5007 + 4959\AA)\slash $I$(H$\beta$)
for the~MPFS field M24.
The contours indicate the H$\alpha$ intensity.

\begin{figure*}
\center{\includegraphics[width=0.8\linewidth]{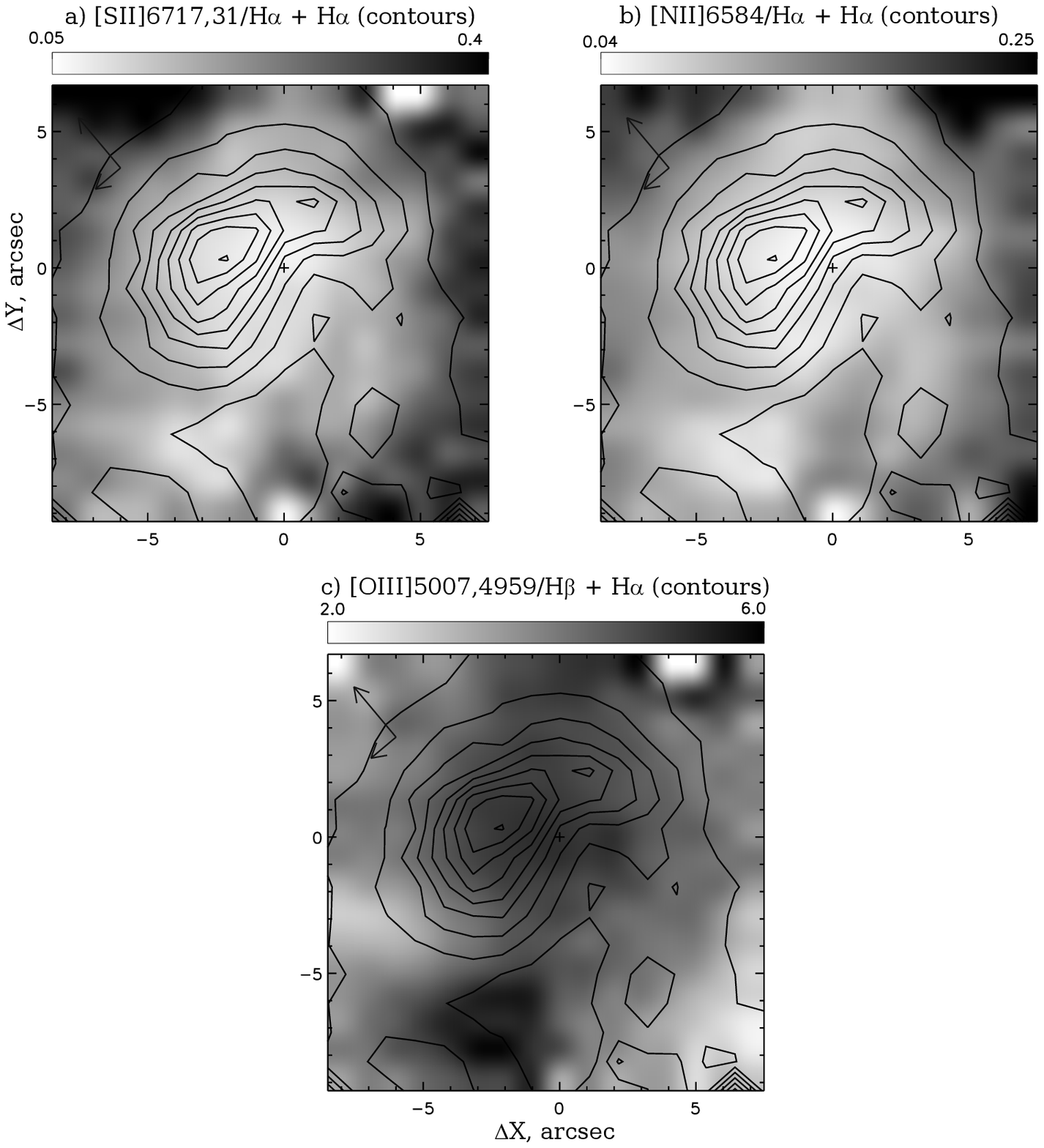}}
\caption{Distribution maps of relative line intensities:
$I$([SII]$\lambda6717+6731$\AA)/$I$(H$\alpha$)~(a),
$I$([NII]$\lambda6584$\AA)/$I$(H$\alpha$)~(b), and
$I$([OIII]$\lambda$5007+4959\AA)/$I$(H$\beta$)~(c) for the MPFS
field M24 (the region of the shell nebula HL111). The arrows in the
upper left corner of the panels indicate the direction northward and
eastward. The contours give the H$\alpha$ intensity. \hfill}
\end{figure*}

\section*{DISCUSSION}

Below, we compare the results of our observations of the emission
spectrum for HII~regions of IC~10 with two new works on the
evolutionary modeling of the emission spectrum for the ionized gas
that surrounds a young star cluster as a function of the gas
metallicity and the cluster age and mass (Levesque et~al.~2009;
Mart\'in-Manj\'on et~al.~2009). In both works, the emission spectrum of
HII~regions is modeled for starburst galaxies and the results of the
computations are particularly interesting for our case of a low gas
metallicity in~IC~10.

The parameters needed to compare the results of our observations
with the mentioned model calculations are primarily the gas density
and metallicity in the galaxy's HII~regions.

\subsection*{Estimating the Density of the Gaseous Medium}

We found the electron density in the investigated nebulae of IC~10 from the [SII]
doublet line intensity ratio from our MPFS observations to be within the range $N_{e}
\simeq 20$ -- $ 250$~cm$^{-3}$. In particular, the density reaches $N_{e} \simeq 60$ --
$100$~cm$^{-3}$ in the nebula HL111c representing the brightest part of the shell HL111
and drops to $N_{e} \simeq 20$ -- $50$~cm$^{-3}$ in the fainter regions of HL111. The
results are presented below in Table~5, which gives the nebular-averaged densities
obtained from our MPFS and long-slit spectroscopic observations.

\subsection*{Estimating the Metallicity of the Gaseous Medium}

The oxygen abundance in individual HII~regions of IC~10 found in
Lozinskaya et~al.~(2009) from long-slit spectroscopy lies within the
range $12+\log(\textrm{O}/\textrm{H}) = 7.59$ -- $8.52$; the
mean metallicity of the gaseous medium in the galaxy is
$12+\log(\textrm{O}/\textrm{H})\simeq 8.17 \pm 0.35$ or $Z = (0.18
\pm 0.14) Z_{\odot}$. The oxygen abundance variations in HII~regions
of~IC~10 turned out to be large; we showed that they are
attributable not only to the measurement errors of the line
intensities but also to the real differences in the galaxy's bright
and faint ionized-gas regions.

The new results of our MPFS observations allow the data from
Lozinskaya et~al.~(2009) to be supplemented. It also seems
interesting to use a different metallicity estimation technique
proposed by Pettini and Pagel~(2004) both to interpret our MPFS
observations and to refine our long-slit spectroscopic data.

The oxygen abundance was estimated in Lozinskaya et~al~(2009) from
the relations found by Pilyugin and Thuan~(2005) using the
calibration dependence of $12+\log(\textrm{O}/\textrm{H})$ on the
relative intensities of strong [OII] and [OIII] lines. To estimate
the abundances of oxygen and nitrogen and sulfur ions, we also used
Eqs.~(6) and~(8) from Isotov~et~al.~(2006).

Yet another estimate of the oxygen abundance in HII~regions can be
obtained by the method proposed by Pettini and Pagel~(2004) using
the interstellar-extinction-independent ratio of the relative line
intensities
$$
\frac {I([\textrm{OIII}]5007)/I(\textrm{H}\beta)}{
I([\textrm{NII}]6584)/I(\textrm{H}\alpha)}.
$$

According to this paper, the oxygen abundance is defined by the
relation $$
12+\log(\textrm{O}/\textrm{H}) = 8.73 -
0.32\times\textrm{O}3\textrm{N}2,$$
 where $$\textrm{O}3\textrm{N}2=\log \frac {I([\textrm{OIII}]5007)}{I(\textrm{H}\beta)}-\log \frac{I([\textrm{NII}]6584)}{I(\textrm{H}\alpha)}$$.
In this case, the oxygen abundance is reliably estimated at
$\textrm{O}3\textrm{N}2<1.9$. Note that Bresolin et~al.~(2004) also
confirmed that this relation works well in the range
$12+\log(\textrm{O}/\textrm{H})<8.4$.

\begin{table*}[p!]
%%% Table:4
\caption{Estimates of the O, N$^{+}$, and S$^{+}$ abundances from
our MPFS and long-slit spectroscopic observations}
\center{\begin{tabular}{|l|c|c|c|c|}
\hline \multicolumn{1}{|c|}{Region}&  O3N2 & $12+
\log(\textrm{O}/\textrm{H})$& $12+ \log(\textrm{N}^{+}/\textrm{H})$
&
$12+ \lg(\textrm{S}^{+}/\textrm{H})$\\
\hline
\multicolumn{5}{|c|}{MPFS observations}\\
\hline
HL111c    & $1.79 \pm 0.12$ &  $8.16 \pm 0.04$ & $6.51 \pm 0.10$ & $5.91 \pm 0.13$\\
HL111d    & $1.75 \pm 0.12$ &  $8.17 \pm 0.04$ & $6.56 \pm 0.11$& $6.02 \pm 0.10$\\
HL111e    & $1.56 \pm 0.09$ &  $8.23 \pm 0.03$ & $6.67 \pm 0.08 $& $6.16 \pm 0.13$\\
\hline
HL100    & $0.95 \pm 0.14$ &   $8.43 \pm 0.04$ & $7.21 \pm 0.15$ &$6.89 \pm 0.17$\\
HL111a   & $1.39 \pm 0.18$ &   $8.29 \pm 0.05$ & $6.95 \pm 0.10$ & $6.50 \pm 0.12$\\
\hline
HL106    & $1.16 \pm 0.18$ &   $8.36 \pm 0.05$ & $6.81 \pm 0.33$& $6.43 \pm 0.43$\\
HL106a   & $1.32 \pm 0.13$ &   $8.31 \pm 0.03$ & $6.61 \pm 0.11$ &$6.08 \pm 0.14$\\
\hline
HL97   & $1.46 \pm 0.21$ &   $8.26 \pm 0.06$ & $6.93 \pm 0.28$&$6.54 \pm 0.31$\\
HL98   & $1.19 \pm 0.18$ &   $8.35 \pm 0.05$ &$7.17 \pm 0.41$ &$6.92 \pm 0.42$\\
\hline
HL4   & $1.33 \pm 0.48$ &   $8.30 \pm 0.16$ &  $7.28 \pm 0.41$&$6.24 \pm 0.22$\\
HL6   & $1.71 \pm 0.36$ &   $8.18 \pm 0.12$ & $7.51 \pm 0.22$ &$7.02 \pm 0.12$\\
\hline
\multicolumn{5}{|c|}{Long-slit spectroscopy}\\
\hline
HL111a  & $ 1.24 \pm 0.14 $ & $8.33 \pm 0.04$ & $6.73 \pm 0.13$ & $6.10 \pm 0.03$ \\
HL111b  & $ 1.01 \pm 0.02 $ & $8.41 \pm 0.01$ &  $6.86 \pm 0.02$ & $6.09 \pm 0.01$ \\
\hline
HL111c  & $ 1.71 \pm 0.01 $ & $8.18 \pm 0.01$ &$6.45 \pm 0.10$ & $5.67 \pm 0.09$ \\
HL111d  & $ 1.48 \pm 0.01 $ & $8.26 \pm 0.01$ & $6.70 \pm 0.01$ & $5.95 \pm 0.01$ \\
HL111e  & $ 1.38 \pm 0.01 $ & $8.29 \pm 0.01$ &$6.69 \pm 0.02$ & $5.97 \pm 0.01$ \\
\hline
HL106   & $ 1.44 \pm 0.16 $ & $8.27 \pm 0.05$ &$6.65 \pm 0.23$ & $6.00 \pm 0.07$ \\
\hline
SS      & $ 0.73 \pm 0.28 $ & $8.50 \pm 0.09$ & $7.07 \pm 0.26$ & $6.61 \pm 0.10$ \\
\hline
HL37    & $ 1.83 \pm 0.37 $ & $8.15 \pm 0.12$ & $6.71 \pm 0.34$ & $5.86 \pm 0.05$ \\
\hline
HL45    & $ 2.03 \pm 0.02 $ & $8.08 \pm 0.01$ & $6.13 \pm 0.10$ & $5.74 \pm 0.01$ \\
\hline
HL50    & $ 1.72 \pm 0.25 $ & $8.18 \pm 0.08$ & $6.54 \pm 0.39$ & $5.81 \pm 0.34$ \\
\hline
HL100   & $ 0.96 \pm 0.06 $ & $8.42 \pm 0.02$ &$6.90 \pm 0.06$ & $6.28 \pm 0.02$ \\
\hline
HL89    & $ 0.90 \pm 0.12 $ & $8.44 \pm 0.04$ &$6.93 \pm 0.03$ & $6.29 \pm 0.03$ \\
\hline
WR M10   & $ 0.96 \pm 0.07 $ & $8.42 \pm 0.02$ &$7.02 \pm 0.07$ & $6.45 \pm 0.03$ \\
\hline
HL67    & $ 1.05 \pm 0.14 $ & $8.39 \pm 0.04$ &$6.94 \pm 0.24$ & $6.20 \pm 0.07$ \\
\hline
HL41    & $ 1.10 \pm 0.05 $ & $8.38 \pm 0.02$ & $6.95 \pm 0.12$ & $6.04 \pm 0.04$ \\
\hline
HL36    & $ 1.64 \pm 0.05 $ & $8.20 \pm 0.02$ & $6.70 \pm 0.14$ & $5.72 \pm 0.04$ \\
\hline
HL22    & $ 1.08 \pm 0.03 $ & $8.38 \pm 0.01$ & $6.91 \pm 0.11$ & $5.98 \pm 0.03$ \\
\hline
HL46-48 & $ 0.87 \pm 0.03 $ & $8.45 \pm 0.01$ & $7.00 \pm 0.09$ & $6.27 \pm 0.04$ \\
\hline
\end{tabular}}
\end{table*}

Based on our MPFS observations, we estimated here the oxygen abundance
$12+\log(\textrm{O}/\textrm{H})$ by the above method for the individual nebulae HL111c,
HL111d, and HL111e that form the shell structure of  HL111 around the M24 group of
WR~stars, for the two HII~regions HL111a and HL100 around the Wolf--Rayet WC4 star M12,
and for the neighborhoods of the stars R10 (HL106 and HL106a), R12 (HL97, HL98), and M2
(HL4, HL6). The results are presented in Table~4. Its columns give: (1)--- name of the
HII~region; (2)--- O3N2 parameter; (3), (4), and (5) --- derived relative O, N$^{+}$,
and S$^{+}$abundances, respectively. As follows from the table, the oxygen abundance
$12 + \log(\textrm{O}/\textrm{H})$ for all nebulae varies within the range from~8.15
to~8.43. The minimum value corresponds to the center of current star formation in the
vicinity of~M24, with the exception of the southwestern fragment~HL111e, where is is
slightly higher. In the remaining HII~regions investigated with the MPFS, the oxygen
abundance is slightly higher and, on average, is close to~8.3.

We used the same method by Pettini and Pagel~(2004) to reestimate
the oxygen abundance in HII~regions based on our long-slit
spectroscopy. The results are also presented in Table~4. Comparison
with the results from Lozinskaya et~al.~(2009) shows that this
method allows the scatter of oxygen abundance estimates to be
reduced considerably compared to the method based on strong [OII]
and~[OIII] lines that we used previously.

To estimate the N$^{+}$ and S$^{+}$ abundances, we, as in Lozinskaya et~al.~(2009),
used Eqs.~(6) and~(8) from Isotov et~al.~(2006), which include the extinction-dependent
relative intensities $I$([NII]$\lambda6548+6584$\AA)/$I$(H$\beta$) and
$I$([SII]$\lambda6717+6731$\AA)/$I$(H$\beta$). Therefore, we slightly modified their
reduction technique to eliminate the possible errors related to the uncertainty in the
correction for interstellar extinction during long-slit spectroscopic observations (the
blue and red regions for spectrograms~PA0, PA132, and PA268 were observed on different
nights). In contrast to Lozinskaya et~al.~(2009), we did not correct the observed
ratios for the color excess in the corresponding region but took the ``theoretical''
ratio $I(\textrm{H}\alpha$):$I(\textrm{H}\beta)=2.86:1.00$, which,
according to Aller~(1984), is valid for typical electron density in a star-forming
region of 20--300~cm$^{-3}$ and electron temperature of $\simeq10\,000$~K. The relative
abundances $12+\log(\textrm{N}^{+}/\textrm{H})$ and $12\log(\textrm{S}^{+}/\textrm{H})$
refined in this way are presented in the last two columns of Table~4.

During our MPFS observations, we took the entire spectrum with the
same exposure time and no problems with the extinction estimation
arose.

We emphasize that the possible errors in the correction for
interstellar extinction do not affect the oxygen abundance estimated
by the method of Pettini and Pagel~(2004), which is derived from the
relative intensities of ``neighbouring'' lines.

According to Pettini and Pagel~(2004), the dispersion of the
dependence of the oxygen abundance on the ratio of
$I$[OIII(5007)]/$I$(H$\beta$) to $I$[NII(6584)]/$I$(H$\alpha$) is
$\pm 0.25$~dex. However, comparison of our estimates from the MPFS
and long-slit spectra shows agreement with an accuracy of at least
$\pm 0.05$~dex. Therefore, the observed difference between the
oxygen abundances in different HII~regions of IC~10, about 0.1~dex,
seems real to us.

The oxygen abundance variations in the interstellar medium of~IC~10,
up to several tenths of~dex, were considered in detail by Magrini
and Gon\c{c}alves~(2009), who pointed out the region of the most
constant oxygen abundance $12+\lg(\textrm{O}/\textrm{H})= 8.2$ at a
galactocentric distance of 0.2--0.5~kpc.

Note that for four regions common to our measurements and that by Magrini and
Gon\c{c}al\-ves~(2009), the results are in good agreement. The nebula~HL45, where our
oxygen abundance estimate is appreciably, by almost 0.3~dex, lower, constitutes an
exception. The high neutral oxygen abundance in this region, which may not be
adequately taken into account in the method of Pettini and Pagel~(2004), is probably
responsible for the discrepancy.

Below, we use the density and metallicity of the gaseous medium
found to choose appropriate theoretical diagnostic models.

\subsection*{Comparing the Observations of~IC~10 with Theoretical
Photoionization Models}

Previously (Lozinskaya et~al.~2009), we have already noted that the
diagnostic diagrams of relative line intensities constructed from
our long-slit spectroscopy agree poorly with the photoionization
models for the gas metallicity found in~IC~10, $Z=0.2 Z_{\odot}$. In
particular, these diagnostic diagrams turned out to be in agreement
with the theoretical calculations by Dopita et~al.~(2006) only for a
metallicity from $Z=0.4 Z_{\odot}$ to \mbox{$Z= (1${--}$2) Z_{\odot}$}.
The comparison with the results of Cid Fernandes et~al.~(2007) and
Asari et~al.~(2007), who summarized the gas metallicity estimates
for $\simeq85\ 000$ starburst galaxies, made in Lozinskaya
et~al.~(2009) also suggests that the HII~regions in~IC~10 agree
excellently with the observations of other galaxies, but only in the
range of metallicities from $Z=0.3 Z_{\odot}$ to $Z= 0.6 Z_{\odot}$.
Given the systematic shift by 0.2~dex revealed by Asari
et~al.~(2007) between their metallicity estimates and those of
Pilyugin and Thuan~(2005) and Izotov et~al.~(2006) for 177
``common'' objects, the lower limit of the range within which the
nebulae of IC~10 investigated in Lozinskaya et~al.~(2009) fell is
$Z=0.2 Z_{\odot}$. Comparison of our constructed diagnostic diagrams
with the models by Charlot and Longhetti~(2001) also showed that the
HII~regions of IC~10 fell within the metallicity range \mbox{$Z=(0.2${--}$1) Z_{\odot}$} (Lozinskaya et~al.~2009).

The technique of theoretical diagnostic curves by Dopita et~al.~(2006) was improved in
the new paper by Levesque et~al.~(2009), who modeled the emission spectrum of
HII~regions for low-metallicity starburst galaxies. These authors used the Starburst99
evolutionary synthesis code and the Mappings~III photoionization code (see Binette
et~al.~1985; Sutherland and Dopita~1993) modified by Groves et~al.~(2004), in which the
influence of dust on the gas ionization was thoroughly taken into account. The spectral
energy distributions for hot stars were taken according to the non-LTE models by
Pauldrach et~al.~(2001) and Hillier and Miller~(1998) with allowance made for the
stellar wind and metal opacity. Levesque et~al.~(2009) considered the models of both a
starburst and continuous star formation in a medium with a metallicity from $z=0.001$
to $z= 0.04$ and with a density $n(\textrm{H})=100$~cm$^{-3}$. The mass loss by O~stars
was taken in the form $dm/dt \simeq Z^{1/2}$; no correction for the metallicity effect
was applied for WR~stars (in contrast to Mart\'in-Manj\'on et~al.~2009; see below).

\begin{figure*}[t!]
\center{\includegraphics[width=0.9\linewidth]{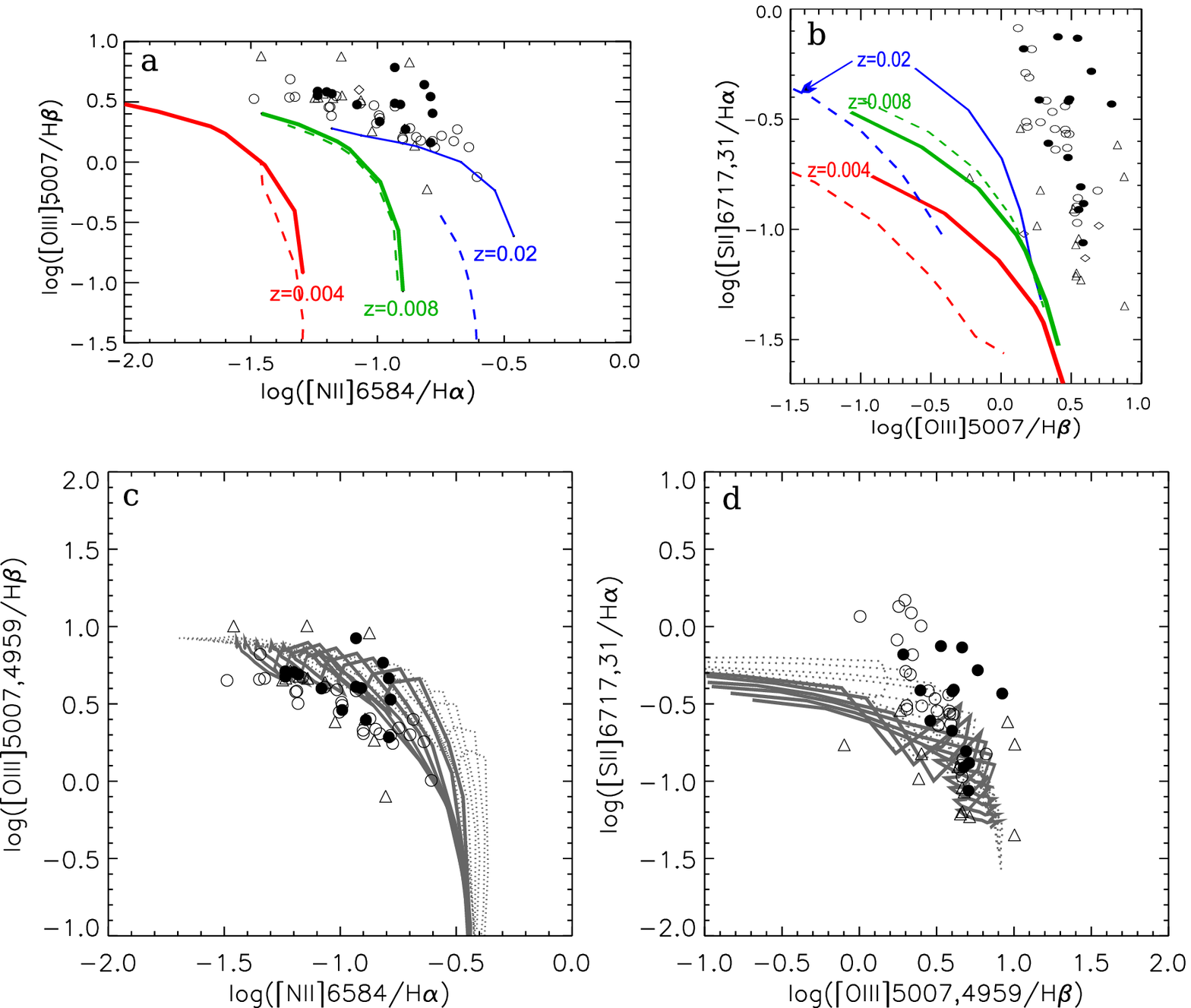}}
\caption{Comparison of the observations of HII~regions in~IC~10
with model calculations. The filled circles, open circles, and
triangles indicate the results of our MPFS observations, our
long-slit spectroscopy, and the data from Magrini and
Gon\c{c}alves~(2009), respectively. Panels~(a) and~(b) show the
$I$([OIII]$\lambda 5007$\AA)/$I$(H$\beta$) vs.
$I$([NII]$\lambda6584$\AA)/$I$(H$\alpha$) and
$I$([SII]$\lambda6717+6731$ \AA)/$I$(H$\alpha$) vs.
$I$([OIII]$\lambda 5007$\AA)/$I$(H$\beta$) diagnostic diagrams
computed by Levesque et~al.~(2009) for a starburst and two ages,
3~Myr (solid curves) and 5~Myr (dashed curves). The computations
are given for three different gas metallicities (shown near the
corresponding curves); $z=0.004$ corresponds to the metallicity of
HII~regions in~IC~10. Panels~(c) and~(d) show the families of
$I$([OIII]$\lambda 4959+5007$\AA)/$I$(H$\beta$) vs.
$I$([NII]$\lambda 6584$\AA)/$I$(H$\alpha$) and
$I$([SII]$\lambda6717+6731$\AA)/$I$(H$\alpha$) vs.
$I$([OIII]$\lambda 4959+5007$\AA)/$I$(H$\beta$) diagnostic
diagrams constructed by Mart\'in-Manj\'on et~al.~(2009) for $z=0.004$.
Different curves correspond to different ionizing-cluster masses
within the range $(0.12${--}$2)\times10^{5} M_{\odot}$ (the lowest
and highest masses are on the right and the left, respectively).
The solid and dashed lines indicate the curves for $N_{e} \simeq
10$~cm$^{-3}$ and $N_{e} \simeq 100$~cm$^{-3}$, respectively.
\hfill}
\end{figure*}

In Figs.~3a and~3b, the results of our MPFS and long-slit spectroscopic observations of
HII~regions and the observational data for IC~10 from Magrini and Gon\c{c}alves~(2009)
are compared with the diagnostic diagrams of relative line intensities, which are
traditionally used to compare observations with ionization models, computed by Levesque
et~al.~(2009). The calculations of $I$([OIII]$\lambda $5007\AA)/$I$(H$\beta$) vs.
$I$([NII]$\lambda $6584\AA)/$I$(H$\alpha$) and $I$([SII]$\lambda $6717 + 6731\AA)/$I$(H$\alpha$) vs. \linebreak
$I$([OIII]$\lambda $5007\AA)/I(H$\beta$) are shown for a starburst
at three gas metallicities and two ages: $t=3$~Myr and $t=5$~Myr. The mean metallicity
of~IC~10 found above  ($Z= 0.2 Z_{\odot}$) corresponds to $z=0.004$; below, we will
estimate the age adopted for our comparison based on the models by Mart\'in-Manj\'on
et~al.~(2009).

As we see, the observations of nebulae in~IC~10 agree poorly with
the model calculations by Levesque et~al.~(2009) for the metallicity
$Z= 0.2 Z_{\odot}$ found here from our MPFS and long-slit
spectroscopic observations and in Magrini and Gon\c{c}alves~(2009).

Mart\'in-Manj\'on et~al.~(2009) used new evolutionary stellar models (PopStar) from Moll\'a
et~al.~(2009). In contrast to many other previously published models, these authors
took the spectral energy distributions for O, B, and WR~stars from Smith et~al.~(2002),
who computed non-LTE models by taking into account the stellar wind and blanketing in a
revised scale of effective temperatures of O~stars and, which is particularly
important, temperatures of WR~stars. The emission line intensities in HII~regions were
computed using the CLOUDY photoionization code (Ferland et~al.~1998), with the decrease
in the abundances of Na, Al, Si, Ca, Fe, and~Ni due to the presence of dust in the
ionized region having been taken into account. The emission spectrum of an HII~region
during a starburst was computed by Mart\'in-Manj\'on et~al.~(2009) for seven star cluster
masses from $0.12\times 10^{5} M_{\odot}$ to $2.0\times 10^{5} M_{\odot}$, for two
densities of the interstellar medium, 10 and 100~cm$^{-3}$, in the range of cluster
ages from~0.1 to 5.2~Myr, and for five gas metallicities in the range from $z =0.001$
to $z=0.04$.

\begin{figure*}[t!]
\center{~~~~~~~~~~~~~~~~~~(a)\hfill

\includegraphics{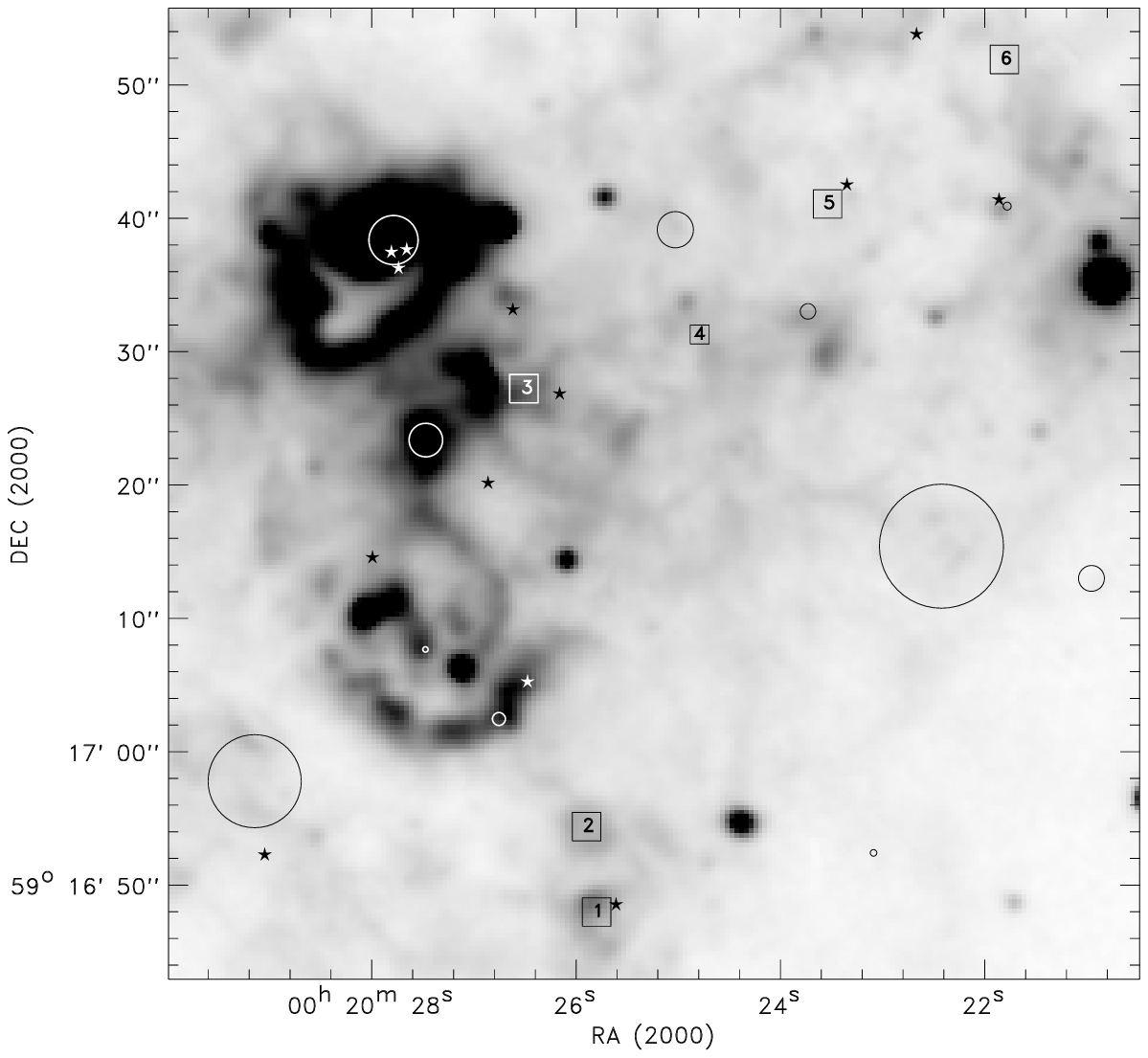}}
\caption{(a) The numbered squares indicate the localization of the
regions where an enhanced intensity
$I$([SII]$\lambda6717+6731$\AA)/$I$(H$\alpha$) compared to the
photoionization curves was revealed in Fig.~3; (b)~examples of the
H$\alpha$ (upper) and~[SII]$\lambda6717$\AA\ (lower) line profiles
from our FPI observations (in arbitrary units). The observed profile
and the high-velocity components identified as an excess above the
Voigt wings are shown. The region numbers indicated at the top
correspond to the numbers of the squares in panel~(a) (as in Fig.~1,
the circles and asterisks represent the star clusters and WR~stars).
\hfill}
\end{figure*}

\begin{figure*}
\center{~~(b)\hfill

\includegraphics[width=0.9\linewidth]{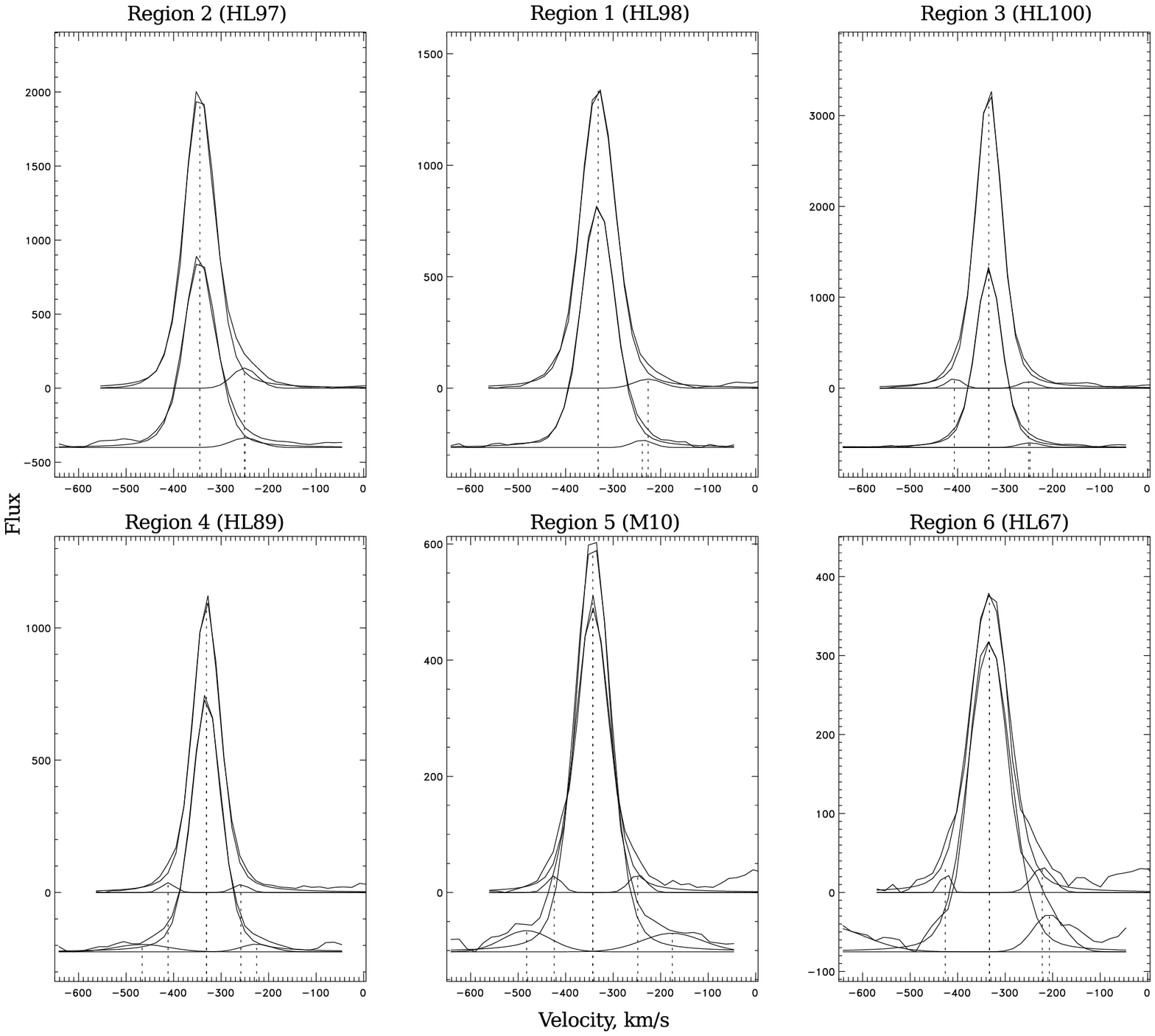}}
\addtocounter{figure}{-1}%
\caption{(Contd.) \hfill}
\end{figure*}

\begin{figure*}
\center{\includegraphics[width=0.95\linewidth]{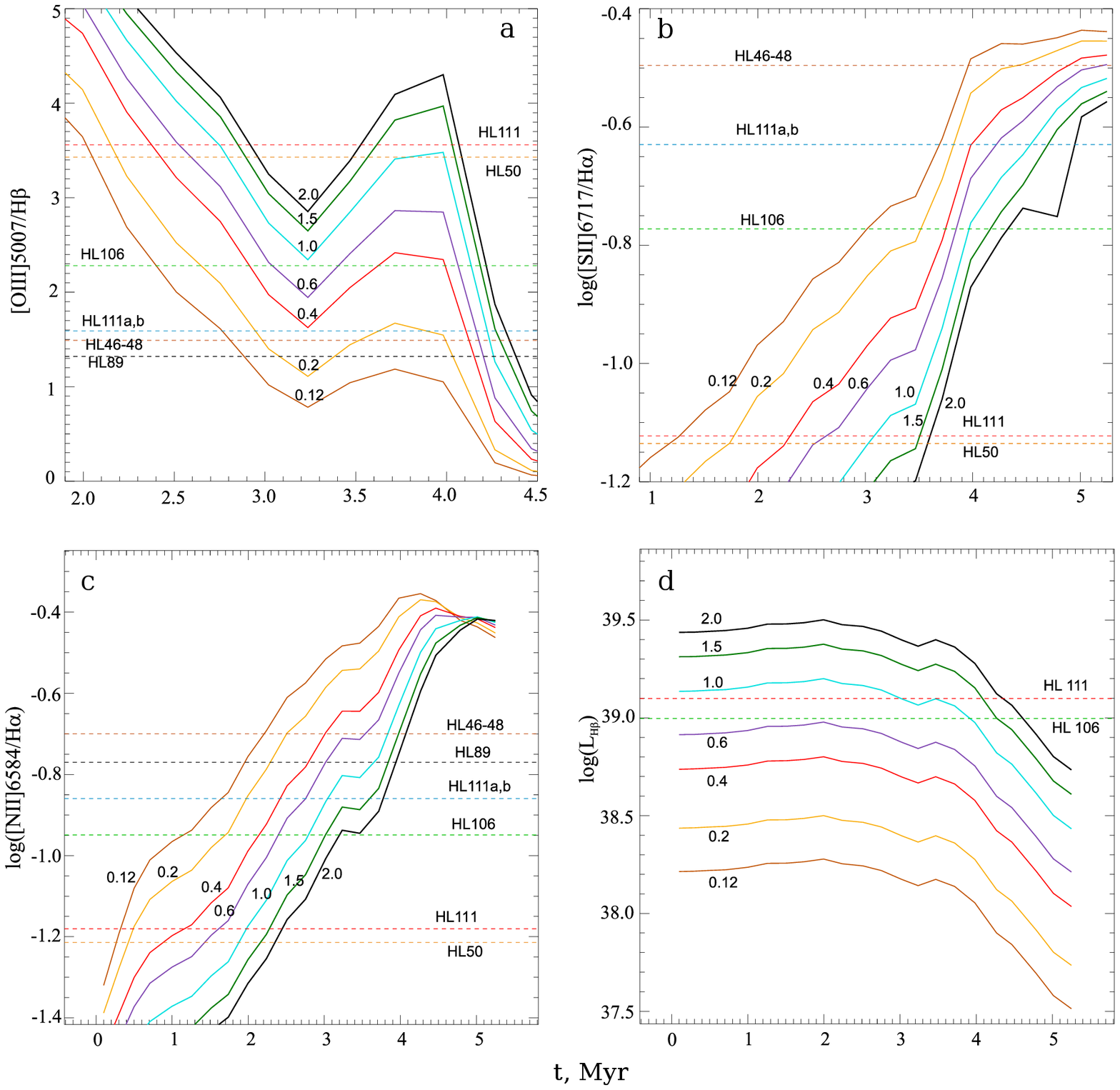}} \caption{Computed dependences
that we constructed from the data of Table~2b presented in the electronic version of
the paper by Mart\'in-Manj\'on et~al.~(2009) for the electron density $N_{e}=100$~cm$^{-3}$
and metallicity $z=0.004$: the age dependences of the line intensities (a)
$I$([OIII]$\lambda 5007$\AA)/$I$(H$_\beta$), (b) $I$([SII]$\lambda
6717$\AA/$I$(H$\alpha$), and (c) $I$([NII]$\lambda6584$\AA)/$I$(H$\alpha$);
(d)~comparison of the observed luminosity $L(\textrm{H}\beta)$ for two HII~regions,
HL111 and HL106, with the computations by Mart\'in-Manj\'on et~al.~(2009). The
corresponding masses of the ionizing clusters are indicated near the model curves.
\hfill}
\end{figure*}

Figures~3c and~3d show the $I$([OIII]$\lambda
4959+5007$\AA)/$I$(H$\beta$) vs. $I$([NII]$\lambda
6584$\AA)/$I$(H$\alpha$) and $I$([SII]$\lambda6717+6731$\AA)/$I$(H$\alpha$) vs. $I$([OIII]$\lambda
4959+5007$\AA)/$I$(H$\beta$) diagnostic diagrams constructed by
Mart\'in-Manj\'on et~al.~(2009) for $z=0.004$. Different curves
correspond to the computations for different ionizing-cluster masses
within the range $(0.12${--}$2)\times10^{5} M_{\odot}$ (the
lowest and highest masses are on the right and the left,
respectively). The solid and dashed lines indicate the curves for
densities $N_{e} \simeq 10$~cm$^{-3}$ and $N_{e} \simeq
100$~cm$^{-3}$, respectively. Different symbols in the figure
indicate the mentioned relative line intensities averaged over the
individual HII~regions of IC~10 that we found from our MPFS and
long-slit spectroscopic observations and the estimates from Magrini
and Gon\c{c}alves~(2009).

As we see, the observations generally agree with the theoretical
dependences for a density within the range $N_{e}\simeq 10
${--}$100$~cm$^{-3}$. Several points in the
$I$([SII]$\lambda6717+6731$\AA)/$I$(H$\alpha$) vs.
$I$([OIII]$\lambda 4959+5007$\AA)/$I$(H$\beta$) diagram deviate
from the theoretical dependence for a photoionized HII~region
toward higher sulfur line intensities. These points correspond to
the long-slit spectrograms passing over the synchrotron
superbubble and over several faint regions. The enhanced [SII]
line intensity here may be related to the gas emission behind the
shock front. Indeed, according to Lozinskaya and Moiseev~(2007),
the synchrotron superbubble is the remnant of a hypernova
explosion.

In the regions HL67, HL89, HL97, HL98, HL100 and near the WR~star
M10, where an enhanced relative [SII] line intensity is also
observed, we searched for high-velocity gas motions that would be
indicative of the action of shock waves. For this purpose, we used
our previous observations at the 6-m SAO telescope with the SCORPIO
instrument and a scanning Fabry--Perot interferometer (FPI) in the
H$\alpha$ and~[SII]$\lambda6717$\AA\ lines, whose results are
discussed in detail in Egorov et~al.~(2010).

The results of our search are presented in Fig.~4. Figure~4a shows the localization of
the regions mentioned above, where an enhanced relative intensity of the
[SII]$\lambda6717+6731$\AA\ lines is observed. The H$\alpha$ and~[SII]$\lambda6717$\AA\
line profiles in these regions constructed from our FPI observations are presented in
Fig.~4b. As follows from the figure, in all regions of an enhanced relative sulfur line
intensity, weak features are actually noticeable in the red and blue line wings at
velocities from~70 to 160~km~s$^{-1}$ relative to the velocity of the peak. In the
[SII]$\lambda6717$\AA\ line, prominent features in the wings are identified at the same
velocities as those in H$\alpha$; the close coincidence of the velocities in the two
lines is a strong argument for the reality of the weak features.

The detected high-velocity motions suggest the action of shock
waves, which can be responsible for the observed enhanced relative
intensity of the [SII] lines.

Thus, we made sure that the results of our observations agree well
with the model calculations by Mart\'in-Manj\'on et~al.~(2009).
Therefore, below we use the theoretical calculations from this paper
to analyze the observational data.

Modeling the emission spectra of HII~regions allows such ionizing-cluster parameters as
the age and mass as well as the ionization parameter to be estimated. For this purpose,
we constructed the corresponding computed dependences using the data from Table~2b in
the electronic version of the paper by Mart\'in-Manj\'on et~al.~(2009). We constructed all
of the computed dependences for the mean electron density $N_{e}=100$~cm$^{-3}$ and
metallicity $z=0.004$ found above. The results are presented in Figs.~5 and~6.

To estimate the ages of the most reliably identified ionization
sources in several HII~regions, we used the computed age dependences
of the line intensities  $I$([OIII]$\lambda 5007$\AA)/$I$(H$\beta$) (Fig.~5a), $I$([SII]$\lambda
6717$\AA/$I$(H$\alpha$) (Fig.~5b), and
$I$([NII]$\lambda6584$\AA)/$I$(H$\alpha$) (Fig.~5c). As a
parameter for the construction of each curve, we used the mass of
the ionizing star cluster, which changes from one curve to another
in the range from $0.12\times 10^{5}$ to $2\times 10^{5} M_{\odot}$.

As follows from the figure, all of the clusters considered --- T54 (in the HII~shell
HL111), T32 (ionizing HL50), T52 (ionizing the region around the star M12), T34?, T24
and T27 (ionizing the regions HL46 and HL48), T47 and poor T43? (ionizing HL89), and
T50 and T53 (probable ionization sources of HL106) --- have ages within the range
from~2.5 to 5~Myr (in what follows, the sign ``?'' in Table~5 marks the clusters that
are possible ionization sources of the corresponding HII~region.) The clusters ionizing
the shell HL111 and the region~HL50 are youngest.

The use of three cluster age indicators based on different spectral lines in Fig.~5
shows certain discrepancies between them and, often, ambiguity. The difficulty of our
estimations is compounded by the fact that all dependences are two-parameter ones (on
cluster age and mass). Therefore, in determining the age, we chose the ranges of
``intersection'' of our estimates according to three criteria, giving preference to the
estimate based on the oxygen lines.

As an example, we will specify the corresponding parameters for the shell HL111.
Judging by Fig.~5, the ages and masses of the clusters ionizing it lie within the
following ranges:
\begin{itemize}
\item from the [OIII]5007/H$\beta$ ratio: \\
  $t=2.1${--}$2.9$~Myr or $t=3.6${--}$4.1$~Myr,\\
$M>1.0 \times10^{5} M_{\odot}$;
\item from the [SII]6717/H$\alpha$ ratio:\\ $t=1.3${--}$3.6$~Myr,\\
$M=(0.12${--}$2)\times 10^{5} M_{\odot}$;
\newpage
\item from the [NII]6584/H$\alpha$ ratio:\\ $t=0.3${--}$2.4$~Myr,\\
$M=(0.12${--}$2)\times 10^{5} M_{\odot}$;\\
\item from the luminosity $L(\textrm{H}\beta)$:\\ $t=3.0${--}$4.3$~
Myr,\\ $M=(1.0${--}$2.0)\times 10^{5} M_{\odot}$ (see below).

\end{itemize}

Our estimate from the relative line intensity [NII] 6584/H$\alpha$
is least reliable, because the nitrogen lines in the nebula HL111
are weak, see \mbox{Table~3}.

Given these discrepancies, we can actually estimate the age only
with an accuracy of $\pm$2~Myr. The scatter of cluster age
estimates may be partly attributable to slight differences in
metallicity and density of the HII~regions associated with them
(we neglected the possible slight differences when estimating the
age).

However, the choice of an age from its possible options is facilitated considerably
when Wolf--Rayet stars are presented in HII~regions. This allows the lower age limit to
be determined. Indeed, at $z=0.004$, according to the evolutionary synthesis models by
Moll\'a et~al.~(2009) used as a basis by Mart\'in-Manj\'on et~al.~(2009), the stage at which
the first Wolf--Rayet nitrogen-sequence (WN) stars appear corresponds to an age of
about 3.2~Myr; the carbon-sequence (WC) stars emerge 1~Myr later. Since the components
of the object M24 in the region HL111c probably belong to the nitrogen sequence of
Wolf--Rayet stars (Vacca et~al.~2007), we can exclude the ionizing-cluster age younger
than 3~Myr.

Since the WC7-type star~R10 is localized near the part of the region
HL106 we observed, we can assign an age older than 4~Myr but no
greater than 5~Myr to the ionizing cluster if we rely on the
evolutionary synthesis models of the stellar population from Moll\'a
et~al.~(2009).

Note also that we find the cluster age ``at the time of ionization'' from the spectra
of HII~regions. The ionization equilibrium time in an evolving HII~region depends on
the metallicity and density of the medium. According to Mart\'in-Manj\'on et~al.~(2009), it
is about 0.2~Myr at $z=0.004$ and $N_{e}=10$~cm$^{-3}$ and decreases with increasing
density. Thus, at $z \geq 0.004$ and the gas density in~IC~10 found above, this time is
considerably (by an order of magnitude!) smaller than the cluster age. Therefore, the
ages of the ionizing cluster and the ionized HII~region are almost equal in our case.

\begin{figure*}
\center{\includegraphics{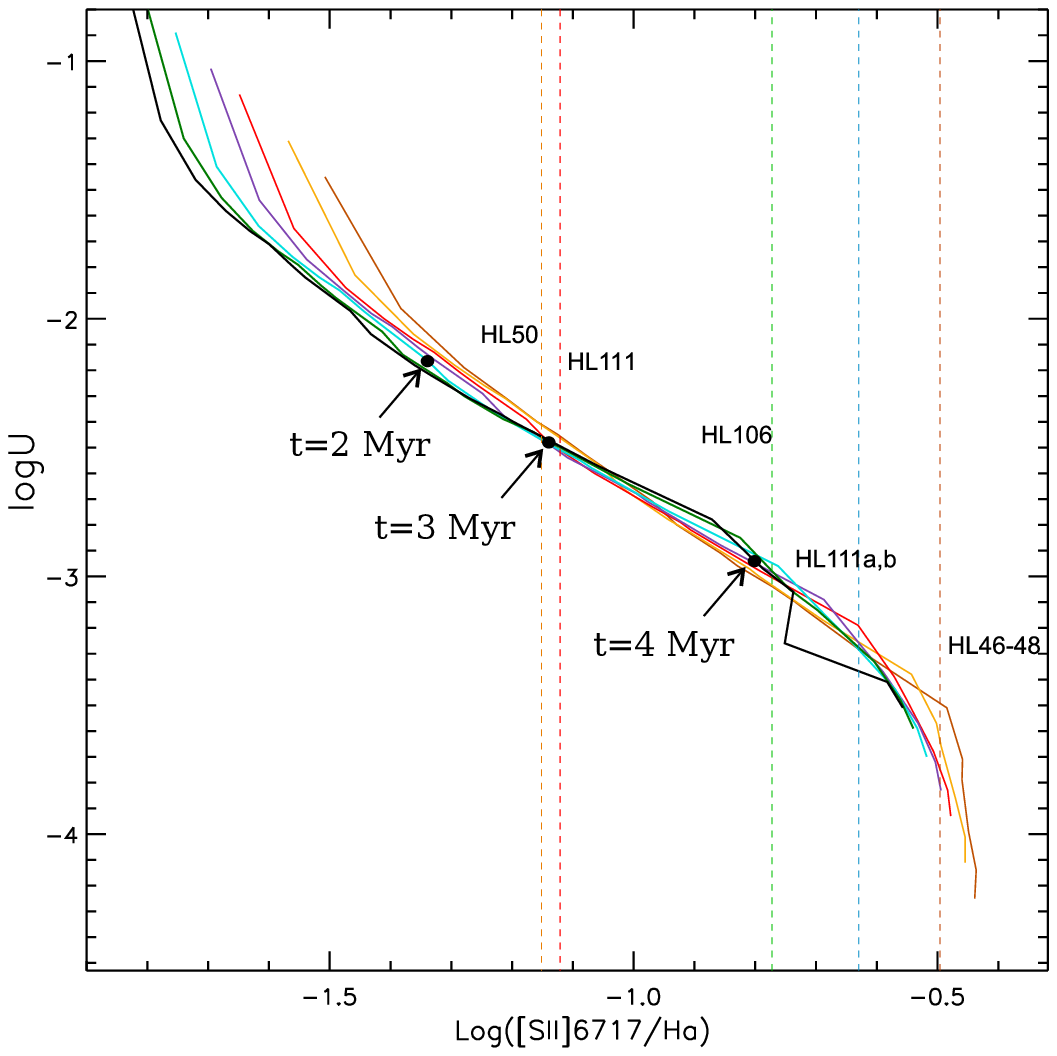}} \caption{Relations between the
line intensity ratio $I$([SII]$\lambda 6717$\AA)/$I$(H$\alpha$)
and ionization parameter~$U$ constructed from the data of Table~2b
in the electronic version of the paper by Mart\'in-Manj\'on
et~al.~(2009) for the metallicity~$z=0.004$ and density $N_{e}$ =
100~cm$^{-3}$. The cluster age changes along the curves from the
upper left corner to the lower right one. The dashed vertical
lines indicate the observed relative intensities corresponding to
the HII~regions. \hfill}
\end{figure*}

Following the recommendations by Mart\'in-Manj\'on et~al.~(2009)
regarding the technique and sequence of parameter determination, we
estimated the ages of the ionizing clusters from the relative line
intensities [OIII]/H$\beta$,~[SII]/H$\alpha$, and [NII]/H$\alpha$ in
the spectra of the surrounding HII~regions. In doing so, we used
WR~stars to refine the lower age limit. The cluster mass can then be
found using the computed H$\beta$ luminosities of the HII~region as
a function of the age for clusters of different masses by comparing
them with the observed luminosities in hydrogen lines. When
estimating the H$\beta$ luminosity, we performed integration over
the area of the corresponding HII~region in the MPFS field. The
results are shown in Fig.~5. We constructed different curves in the
figure from the data of Table~2b in the electronic version of the
paper by Mart\'in-Manj\'on et~al.~(2009) for clusters of different
masses. The horizontal lines indicate the luminosities of two
nebulae found from our MPFS observations and corrected for
extinction: HL111 around the cluster~T54, which includes~M24 (Hunter
4--1 and 4--2), and HL106, which is most likely ionized by the
clusters~T50 and~T53.

We integrated the H$\beta$ emission from the shell~HL111 to estimate its luminosity
$L(\textrm{H}\beta)\simeq 1.3\times 10^{39} $~erg~s$^{-1}$ almost over the entire
nebula covered by the MPFS field, with the exception of its small area on the east. The
age of the cluster~T54, about 3--4~Myr, can be reliably estimated both from the
relative line intensities mentioned above and from the color indices of stars in
Hunter~(2001) and Vacca et~al.~(2007). Accordingly, we find the mass of the cluster T54
to be $M\simeq 1.0\times 10^{5} M_{\odot}$.

The mass estimate for the clusters ionizing~HL106 is less reliable. First, only half of
the nebula falls within the MPFS field and the luminosity $L(\textrm{H}\beta) =
1.0\times 10^{39}$~erg~s$^{-1}$ found has a low accuracy. Second, the age estimates
from the relative line intensities $I$([OIII]/H$\beta$) and $I$([NII]/H$\alpha$) (about
4~Myr) and in Hunter~(2001) (20--30~Myr) disagree significantly. Besides, we showed
previously (Egorov et~al.~2010) that the nebula HL106 could be an ionized shell
surrounding a dense CO~cloud. Therefore, the mass estimate for the clusters T50 and
T53, $ M\simeq 0.5\times 10^{5} M_{\odot}$, is only approximate.

When estimating the cluster masses, we also took into account the
positions of the HII~regions in Fig.~5, where each curve corresponds
to an age of the ionizing star cluster in the range from \mbox{$0.12\times
10^{5}$} to $2\times 10^{5} M_{\odot}$.

\sloppy{The computations by Mart\'in-Manj\'on et~al.~(2009) also allow the ionization
parameter~($U$) to be estimated from the relative line intensity $I$([SII]$\lambda
6717$\AA)/$I$(H$\alpha$). Figure~6 shows the relations between the relative line
intensity $I$([SII]$\lambda 6717$\AA)/$I$(H$\alpha$) and parameter~$U$ that we
constructed from the data of Table~2b in the electronic version of the above paper for
the gas metallicity in this galaxy $z=0.004$ and density $N_{e} = 100$~cm$^{-3}$. As we
see, the relation between these parameters is virtually independent of the cluster
mass. The age in Fig.~6 rises along the merging curves from the upper left corner to
the lower right one. The vertical lines indicate the values of $I$([SII]$\lambda
6717$\AA)/$I$(H$\alpha$) that we found in the investigated HII~regions.
}

\begin{table*}[t!]
%%% Table:5
\caption{Parameters of the HII~regions and the clusters ionizing
them}
\center{\begin{tabular}{|l|l|c|c|c|c|c|}
\hline \multicolumn{1}{|c|}{Region}&
\multicolumn{1}{c|}{Cluster} &  $N_{e}$, cm$^{-3}$ &
$L$(H$\beta$), $10^{39}$ erg~s$^{-1}$& Age, Myr&
Mass, $10^{5} M_{\odot}$& $\lg(U)$\\
\hline
HL111 & T54       & \phantom{0}70   & 1.3  & 3.5--4.0 & 1.0--1.5 &--2.5\phantom{0} \\
\hline
HL106  & T50, T53  & 200  & 1.0  & 4.0--4.5 & ${\ge }1.0$  & --3.0\phantom{0}\\
\hline
HL50   & T32 & \phantom{0}30   & -- & 3.4--3.6 or & 1.5--2.0 or& --2.47 \\
       & & &            & 2.5--3.0 & 0.4--1.0  & \\
\hline
HL111a,b & T52   & 200  & -- & 3.8--4.5 & 0.2--0.6 & --3.27 \\
HL100 & & & & & & \\
\hline
HL46-48 & T34?, T24, T27&  --  &  --  & 4.0--5.0 & 0.2--0.4& --3.6\phantom{0}\\
\hline
HL89  & T47, T43? & 200 & -- & ${>}3$ & ${\ge }0.2$ & --\\
\hline
\end{tabular}}
\end{table*}

For the arc HL111c near~M24, Vacca et~al.~(2007) found the flux of ionizing radiation
with a wavelength shorter than 912~\AA \ to be $Q(\textrm{H}) = (3${--}$6)\times
10^{50}$~phot.~s$^{-1}$. Our estimate of the photon luminosity for the entire shell
HL111 (HL111c+HL111d+HL111e) from the observed H$\beta$ flux corrected for
interstellar extinction is $Q(\textrm{H})\geq3.2\times 10^{51}$~phot.~s$^{-1}$.

For a cluster age of 3--4~Myr, Moll\'a et~al.~(2009) give the ratio of the photon
luminosity of the ionizing stars to the cluster mass $\log(Q(\textrm{H})/M) = 46.75$ at
$z = 0.004$. If the cluster mass in HL111 is taken to be $M \simeq 10^{5} M_{\odot}$,
then we find $Q(\textrm{H})= 5.6\times 10^{51}$~phot.~s$^{-1}$. Thus, the estimates of
the flux of ionizing radiation obtained from the observations and model calculations
are in good agreement.

Table~5 presents all of the parameters of the clusters ionizing the most thoroughly
studied HII~regions obtained in this paper. Its respective columns give: (1)--- name of
the nebula, (2)--- ionizing cluster according to the list by Tikhonov and
Galazutdinova~(2010) (the sign ``?'' marks the clusters whose contribution is also
possible), (3)--- gas density in the nebula, (4)--- line luminosity
$L(\textrm{H}\beta$), (5)--- cluster age, (6)--- cluster mass, (7)--- ionization
parameter.

Summarizing the results of our analysis of the gas emission spectrum and ionization
sources in~IC~10 based on our long-slit and MPFS observations at the 6-m SAO telescope,
we  conclude the following.

Our new estimates of the gas metallicity in the galaxy from both
series of observations using the technique by Pettini and
Pagel~(2004) different from that in Lozinskaya et~al.~(2009) allowed
the scatter of values for different nebulae to be reduced
considerably, but they did not change the galaxy-averaged value of
$Z= 0.2 Z_{\odot}$ that we found previously. The new estimates of
the relative abundances of oxygen and nitrogen and sulfur ions in
nebulae from our MPFS observations and refined estimates from
long-slit spectrograms are gathered in Table~4.

In Lozinskaya et~al.~(2009), we showed that the results of our observations of
HII~regions in~IC~10 agree poorly with the previously published model calculations for
photoionized nebulae for the metallicity found. Comparison with the new improved
ionization models by Levesque et~al.~(2009) showed that the observed $I$([OIII]$\lambda
5007$\AA)/$I$(H$\beta$) vs. $I$([NII]$\lambda6584$\AA)/$I$(H$\alpha$) and vs.
$I$([SII]$\lambda6717+6731$\AA)/$I$(H$\alpha$) diagnostic diagrams agree poorly with
the computations for the gas metallicity in the galaxy that we and Magrini and
Gon\c{c}alves~(2009) found.

At the same time, comparison of the observed diagnostic diagrams of relative line
intensities with the new ionization models by Mart\'in-Manj\'on et~al.~(2009) showed that
the observations of~IC~10 are in good agreement with these computations. The effect of
an enhanced relative intensity $I$([SII])/$I$(H$\alpha$) in several nebulae compared to
the model photoionized HII~region that is clearly seen in Fig.~3 may be attributable to
the contribution from the gas emission behind the front of the shocks triggered by
supernova explosions and stellar winds. Indeed, some of the ``outlying'' points
correspond to the synchrotron superbubble --- the remnant of a hypernova explosion (see
Lozinskaya and Moiseev~2007). In other nebulae, where an enhancement of the relative
intensity $I$([SII])/$I$(H$\alpha$) is also observed, our FPI observations revealed
weak features in the H$\alpha$ and~[SII] line profiles at supersonic velocities as high
as 100--150~km~s$^{-1}$ relative to the velocity of the peak, which is indicative of
the action of shock waves. A prominent action of shock waves in~IC~10 is also suggested
by the filamentary structure of the entire H$\alpha$ emission region in the galaxy.

Thus, we made sure that among the large number of published theoretical evolutionary
models for the emission spectrum of HII~regions during a starburst, our observations of
the galaxy~IC~10 are in best agreement with the computations by Mart\'in-Manj\'on
et~al.~(2009). Therefore, the presented results of our MPFS observations and our
long-slit spectroscopic data are interpreted here in terms of the improved ionization
models by these authors. Using these computations, we determined the ionization
parameter and the ages and masses of the ionizing clusters from the spectra of several
most thoroughly studied HII~regions in the galaxy. When estimating the age, we also
took into account the presence of nitrogen- and carbon-sequence WR stars , which gives
its lower limit. The cluster ages and masses found in the investigated star-forming
region of IC~10 lie within the range 2.5--5~Myr and in the range from $0.2 \times
10^{5}$ to~$10^{5} M_{\odot}$.

\section*{CONCLUSIONS}

This paper continues the studies of the ionized gas emission spectrum in the Irr
starburst~galaxy IC~10 begun by Lozinskaya et~al.~(2009) based on long-slit
spectroscopy. Here, we presented new results of our observations of selected galactic
fields with the panoramic~MPFS; we also used the results of our observations with a
long-slit spectrograph and a scanning Fabry--Perot interferometer.

To estimate the gas metallicity in the galaxy, we used the method by Pettini and
Pagel~(2000) different from that used in Lozinskaya et~al.~(2009). The gas metallicity
was determined by this method both from our new MPFS observations and from our
long-slit spectroscopy. The metallicity estimated in this way is characterized by a
smaller scatter of values for individual nebulae but it does not changes the
galaxy-averaged value.

Previously (Lozinskaya et~al.~2009), we showed that the results of
our observations of HII~regions in~IC~10 agree poorly with the
published model calculations for photoionized nebulae at the
metallicity $Z= 0.2 Z_{\odot}$ found. In this paper, we made such a
comparison with two new computed ionization models from Levessque
et~al.~(2009) and Mart\'in-Manj\'on et~al.~(2009).

We showed that the observations of~IC~10 agree well with the computations by
Mart\'in-Manj\'on et~al.~(2009). We explained the noticeable (in several nebulae) effect of
an enhanced relative intensity $I$([SII])/$I$(H$\alpha$) compared to the model
photoionized HII~region by the contribution from the gas emission behind the fronts of
the shocks triggered by supernova explosions and stellar winds. Indeed, some of the
``outlying'' points correspond to the synchrotron superbubble --- the remnant of a
hypernova explosion (Lozinskaya and Moiseev~2007); in other nebulae with an enhanced
relative intensity $I$([SII])/$I$(H$\alpha$), our FPI observation revealed weak
features in the H$\alpha$ and~[SII] line profiles at supersonic speeds as high as
100--150~km~s$^{-1}$, suggesting the action of shocks.

Having made sure that among the large number of published theoretical evolutionary
models for the emission spectrum of HII~regions in a starburst galaxy, our observations
of IC~10 agree best with the computations by Mart\'in-Manj\'on et~al.~(2009), we used these
computations to analyze our MPFS and long-slit spectroscopic observations. Based on the
ionization models of these authors, we determined the ionization parameter and the ages
and masses of the ionizing clusters from the spectra of several most thoroughly studied
HII~regions. In our estimations, we took also into account the presence of WR~stars,
which gives a lower limit for the age. The cluster ages found in the investigated
star-forming region of~IC~10 lie within the range 2.5--5~Myr, while the masses of the
ionizing star clusters lie within the range from $0.2 \times10^{5}$ to~$10^{5}
M_{\odot}$.

\bigskip

After our paper was accepted for publication, a paper by
L\'opez-S\'anchez et al. (2010) appeared (ArXiv1010.1806) with the
results of the most detailed spectral observations of the HL111
region with 3.5 m telescope at Calar Alto Observatory with a
spatial resolution of $1''$. For the whole HL111 region these
authors derived the temperature T${_e}$=10500 K, the oxygen
abundance $12+\lg(\textrm{O}/\textrm{H})= 8.26 \pm 0.09$ and the
age of the recent star-formation episode of about 3.3 Myr -- all
the data are in full agreement with our results. The only
discrepancy in the fluxes of hydrogen lines in HL111 may be due to
our not too reliable absolute calibration of H$\beta$ line.
 In two spaxels coinciding with WR star M24B these authors
 tentatively detected the broad HeII 4686 line produced by a
 single WNL star. At the same location a possible N/O and He/H
 medium enrichment was found expected for the pollution by the
 WR-ejecta.

\section*{ACKNOWLEDGMENTS}

This work was supported by the Russian Foundation for Basic
Research (project nos.~07-02-00227 and 10-02-00091). We thank
N.A.~Tikhonov, M.E.~Sharina and O.A.~Galazutdinova for their help
in discussing the accuracy of the coordinates of star clusters.
O.V.~Egorov and A.V.~Moiseev thank the hank the `Dynasty' Fund for
financial support. The work is based on the observational data
obtained with the 6-m SAO telescope funded by the Ministry of
Science of Russia (registration no. 01-43). When working on the
paper, we used the NASA/IPAC Extragalactic Database (NED) operated
by the Jet Propulsion Laboratory of the California Institute of
Technology under contract with the National Aeronautics and Space
Administration (USA).

\bigskip
Translated by V.~Astakhov

\bigskip

\centerline{\bf{REFERENCES} }
\bigskip

\begin{enumerate}

\item V.~L.~Afanasiev, S.~N.~Dodonov, and A.~V.~Moiseev, \emph{Stellar Dynamics: from Classic to Modern,} Ed. by L.~P.~Osipkov and I.~I.~Nikiforov (SPb. Univ., St.-Petersburg), p.103 (2001)   .

\item V.~L.~Afanasiev and A.~V.~Moiseev, Astron. Lett. \textbf{31}, 194 (2005).

\item L.~H.~Aller, \emph{Physics of Thermal Gaseous Nebulae} (Reidel, Dordrecht, 1984).

\item N.~V.~Asari, R.~Cid Fernandes, G.~Stasinska, et al., \hyphenation{MNRAS} \textbf{381}, 263 (2007).

\item L.~Binette, M.~A.~Dopita, and I.~R.~Tuohy, Astrophys. J. \textbf{297}, 476 (1985).

\item F.~Bresolin, D.~R.~Garnett, and R.~C.~Kennicutt, Jr., Astrophys. J. \textbf{615}, 228 (2004).

\item A.~Bullejos and M.~Rozado, Rev. Mex. (Ser. de Conf.) \textbf{12}, 254 (2002).

\item S.~Charlot and M.~Longhetti, Mon. Not. R.~Astron. Soc. \textbf{323}, 887 (2001).

\item K.~T.~Chyzy, J.~Knapik, D.~J.~Bomans, et al., Astron. Astrophys. \textbf{405}, 513 (2003).

\item R.~Cid Fernandes, N.~V.~Asari, L.~Sodre, et al., MNRAS \textbf{375}, 16 (2007).

\item P.~A.~Crowther, L.~Drissen, J.~B.~Abbott, et al., Astron. Astrophys. \textbf{404}, 483 (2003).

\item M.~A.~Dopita, J.~Fischera, R.~S.~Sutherland, et al., Astrophys. J.~Suppl. Ser. \textbf{167}, 177 (2006).

\item O.~V.~Egorov, T.~A.~Lozinskaya, and A.~V.~Moiseev, Astron. Rep. \textbf{54}, 277 (2010).

\item G.~I.~Ferland, K.~T.~Korista, D.~A.~Verner, et al., Publ. Astron. Soc. Pacif. \textbf{110}, 761 (1998).

\item A.~Gil de Paz, B.~F.~Madore, and O.~Pevunova, Astrophys. J.~Suppl. Ser. \textbf{147}, 29 (2003).

\item B.~Groves, M.~Dopita, and R.~Sutherland, Astroph. J.~Suppl.Ser. \textbf{153}, 9 (2004).

\item D.~J.~Hillier and D.~L.~Miller, Astrophys. J. \textbf{496}, 407 (1998).

\item P.~Hodge and M.~G.~Lee, Publ. Astron. Soc. Pacif. \textbf{102}, 26 (1990).

\item D.~A.~Hunter, Astrophys. J. \textbf{559}, 225 (2001).

\item Y.~I.~Izotov, G.~Stasinska, G.~Meynet, et al., Astron. Astrophys. \textbf{448}, 955 (2006).

\item A.~Leroy, A.~Bolatto, F.~Walter, and L.~Blitz, Astrophys. J. \textbf{643}, 825 (2006).

\item E.~M.~Levesque, L.~J.~Kewley, and K.~L.~Larson, astro-ph/0908.0460 (2009).

\item A.~R.~L\'opez-S\'anches, A.~Mesa-Delgado, L.~L\'opez-Mart\'in and
C.~Esteban, astro-ph/1010.1806L (2010)

\item T.~A.~Lozinskaya and A.~V.~Moiseev, MNRAS \textbf{381}, 26L (2007).

\item T.~A.~Lozinskaya, A.~V.~Moiseev, N.~Yu.~Podorvanyuk, and A.~N.~Burenkov, Astron. Lett. \textbf{34}, 217 (2008).

\item T.~A.~Lozinskaya, O.~V.~Egorov, A.~V.~Moiseev, and D.~V.~Bizyaev, Astron. Lett. \textbf{35}, 730 (2009).

\item L.~Magrini and D.~R.~Gon\c{c}alves, MNRAS \textbf{398}, 280 (2009).

\item M.~L.~Mart\'in-Manj\'on, M.~L.~Garc\'ia-Vargas, M.~Moll\'a, and A.~I.~D\'iaz, MNRAS \textbf{403}, 2012 (2010), astro-ph/0912.4730 (2009).

\item P.~Massey and S.~Holmes, Astrophys. J. \textbf{580}, L35 (2002).

\item P.~Massey, T.~E.~Armandroff, and P.~S.~Conti, Astron. J. \textbf{103}, 1159 (1992).

\item P.~Massey, K.~Olsen, P.~Hodge, et al., Astron. J. \textbf{133}, 2393 (2007).

\item M.~Moll\'a, M.~L.~Garc\'ia-Vargas, and A.~Bressan, MNRAS \textbf{398}, 451 (2009).

\item A.~W.~A.~Pauldrach, T.~L.~Hoffmann, and M.~Lennon, Astron. Astrophys. \textbf{375}, 161 (2001).

\item M.~Pettini and B.~E.~J.~Pagel, Mon. Not. R.~Astron. Soc. \textbf{348}, L59 (2004).

\item L.~S.~Pilyugin and T.~X.~Thuan, Astrophys. J. \textbf{631}, 231 (2005).

\item M.~G.~Richer, A.~Bullejos, J.~Borissova, et al., Astron. Astrophys. \textbf{370}, 34 (2001).

\item M.~Rosado, M.~Valdez-Gutierrez, A.~Bullejos, et al., ASP Conf. Ser. \textbf{282}, 50 (2002).

\item P.~Royer, S.~J.~Smartt, J.~Manfroid, and J.~Vreux, Astron. Astrophys. \textbf{366}, L1 (2001).

\item N.~Sanna, G.~Bono, P.~B.~Stetson, et al., Astrophys. J. \textbf{699}, L84 (2009).

\item M.~E.~Sharina, R.~Chandar, T.~H.~Puzia, et al., MNRAS, astro-ph/1002.2144 (2009).

\item L.~Smith, R.~Norris, and P.~Crowther, MNRAS \textbf{337}, 1309 (2002).

\item R.~Sutherland and M.~A.~Dopita, Astrophys. J. \textbf{88}, 253 (1993).

\item J.~C.~Thurow and E.~M.~Wilcots, Astron. J. \textbf{129}, 745 (2005).

\item N.~A.~Tikhonov, O.~A.~Galazutdinova, Astron. Lett. \textbf{35}, 748 (2009).

\item W.~D.~Vacca, C.~D.~Sheehy, and J.~R.~Graham, Astrophys. J. \textbf{662}, 272 (2007).

\item E.~M.~Wilcots and B.~W.~Miller, Astron. J. \textbf{116}, 2363 (1998).

\item H.~Yang and E.~D.~Skillman, Astron. J. \textbf{106}, 1448 (1993).

\item J.~Yin, L.~Magrini, F.~Matteucci, et al., astro-ph/1005.3500 (2010).

\item D.~B.~Zucker, Bull. Am. Astron. Soc. \textbf{32}, 1456 (2000).

\item D.~B.~Zucker, Bull. Am. Astron. Soc. \textbf{34}, 1147 (2002).
\end{enumerate}

\end{document}